\documentclass[prl,aps,twocolumn, floatfix]{revtex4-1}
\usepackage{amssymb}
\usepackage{graphicx}
\usepackage{dcolumn}
\usepackage{bm,amsmath,verbatim}
\usepackage{epstopdf}

\def\be{\begin{equation}}
\def\ee{\end{equation}}
\def\bea{\begin{eqnarray}}
\def\eea{\end{eqnarray}}
\def\bse{\begin{subequations}}
\def\ese{\end{subequations}}

\def\be{\begin{eqnarray}}
\def\ee{\end{eqnarray}}

\newcommand{\ua}{\uparrow}
\newcommand{\da}{\downarrow}

\begin{document}

\title{Probing a topological quantum critical point in semiconductor-superconductor heterostructures}

\author{Sumanta Tewari$^{1}$}
\author{J. D. Sau$^{2}$}
\author{V. W. Scarola$^{3}$}
\author{Chuanwei Zhang$^{4}$}
\author{S. Das Sarma$^2$}

\begin{abstract}
Quantum ground states on the non-trivial side of a topological quantum critical point (TQCP) have unique properties that make them attractive candidates for quantum information applications. A recent example is provided by $s$-wave superconductivity on a semiconductor platform, which is tuned through a TQCP to a topological superconducting (TS) state by an external Zeeman field. Despite many attractive features of TS states, TQCPs themselves do not break any symmetries, making it impossible to distinguish the TS state from a regular superconductor in conventional bulk measurements. Here we show that for the semiconductor TQCP this problem can be overcome by tracking suitable bulk transport properties across the topological quantum critical regime itself.  The universal low-energy effective theory and the scaling form of the relevant susceptibilities also provide a useful theoretical framework in which to understand the topological transitions in semiconductor heterostructures. Based on our theory, specific bulk measurements are proposed here in order to characterize the novel TQCP in semiconductor heterostructures.
\end{abstract}
\affiliation{$^{1}$Department of Physics and Astronomy, Clemson University, Clemson, SC
29634 USA\\$^2$Condensed Matter Theory Center and Joint Quantum Institute, Department of Physics, University of Maryland, College Park, Maryland, 20742-4111, USA\\$^3$Department of Physics, Virginia Tech, Blacksburg, VA 24061, USA\\$^{4}$Department of Physics and Astronomy, Washington State University,
Pullman, WA 99164 USA}
\maketitle

\begin{figure}
\centering
\includegraphics[scale=0.4,angle=0]{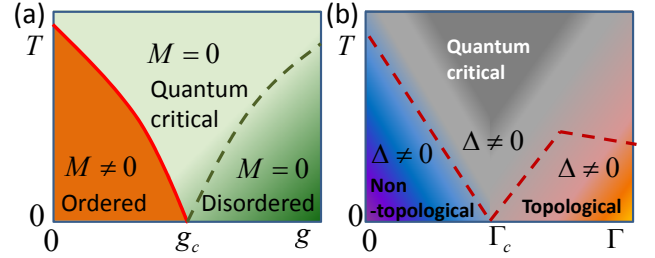}
\caption{(a) Phase diagram associated with a conventional QCP in the ($T-g$)-plane, where $g$ is the tuning parameter in the Hamiltonian. Solid curves denote true phase transitions, while dashed curves denote
only a crossover.
(b) Finite-temperature phase diagram associated with the TQCP in a
 spin-orbit coupled semiconductor.
The tuning parameter $\Gamma$  represents a suitably directed Zeeman splitting. The superconducting pair potential $\Delta$ is
perfectly continuous and non-zero at, and on either side of, the TQCP ($T=0,\Gamma=\Gamma_c$). Consequently, all lines
on this diagram are crossover lines defined by $k_B T = E_0$ where $E_0$ is the zero-temperature energy gap.
}
\label{fig:phase_diagram}
\end{figure}

\textbf{Introduction:}
 Quantum critical points (QCP) separate two many-body quantum
ground states distinguishable by a macroscopic order parameter
$M$ (see Fig.~\ref{fig:phase_diagram}a) \cite{Sachdev}.
In Fig.~\ref{fig:phase_diagram}a
the solid curve denotes a true phase transition line while the dashed curve, given by $k_BT \sim E_0$ where $E_0$
 is the zero-temperature energy gap, represents only a crossover. The two curves meet at the QCP at a specific value of the tuning parameter
 $g=g_c$, straddling a
finite regime in the ($T-g$) plane usually called the quantum critical
(QC) regime \cite{CHN}. Quite interestingly, at the QC regime the effects of the
zero-$T$ quantum phenomena and the associated QCP are manifest even at finite temperatures. This so-called `quantum
fan' region, where quantum criticality manifests far from $g=g_c$ at finite temperatures, enables experimental studies
of QCP which are strictly speaking $T=0$ phase transitions tuned by the parameter $g$.

\emph{Topological} quantum critical points also separate, based on distinct topological properties, two macroscopic
quantum ground states, although the states in question now have exactly the same symmetries and thus
cannot be distinguished by any local order
parameter or bulk measurements (Fig.~\ref{fig:phase_diagram}b) \cite{Wen-Book}.
Frequently, the quantum state on the topological side of a TQCP can be
distinguished by certain non-trivial statistical properties of its
excitations \cite{Kitaev-2003,Read,Ivanov,Nayak-RMP}, as well as a novel ground state quantum degeneracy which is not associated with any symmetry in the underlying Hamiltonian \cite{Nayak-RMP}. An example is provided by an electron- or hole-doped semiconductor
thin film or nanowire with $s$-wave superconductivity tuned through a TQCP by an externally applied Zeeman splitting $\Gamma$.
This system has recently been studied
extensively after it was pointed out by Sau \emph{et al.} \cite{Sau} that for $\Gamma$ greater than a critical value $\Gamma_c$ this system supports novel non-Abelian topological states \cite{Index-Theorem,Alicea-Tunable,Roman,Oreg,Sau-PRB,Alicea-Network,Hassler,Sau-Tewari-Universal,Akhmerov-Conductance,Mao,Sau-Interferometry,Sumanta-NJP}.
For $\Gamma > \Gamma_c$, defects in the (proximity-induced) $s$-wave
pair potential $\Delta $ can support localized topological zero-energy
excitations called Majorana fermions. Majorana fermions,
with second-quantized operators $\gamma$ satisfying $\gamma^{\dagger }=\gamma$, follow non-Abelian exchange statistics under
pair-wise exchange of the coordinates \cite{Kitaev-2003,Read,Ivanov,Nayak-RMP}. Majorana fermions have been
predicted to be useful for building a topological quantum computer which is
intrinsically fault-tolerant to all local environmental decoherence \cite{Kitaev-2003,Nayak-RMP}.

The existence of Majorana fermions in the defects of $\Delta$ notwithstanding,
 $\Delta$ itself remains perfectly continuous and non-zero on
both sides of the TQCP in a semiconductor. This leads to there being no qualitative
difference between the two states in conventional bulk measurements. Thus, it appears that simple transport quantities such as resistance are
unable to demonstrate the emergence of the TS state.
In this paper we propose a very specific (and experimentally simple) scheme for the direct observation of the TQCP
in bulk measurements provided such measurements access the so-called \emph{topological} quantum critical regime, which occurs
\emph{before} the system settles into the TS state at large $\Gamma > \Gamma_c$.
We believe that our work brings topological quantum phase transitions (TQPT) in semiconductors explicitly into the mainstream of quantum critical phenomena, something that was only
implicit in the extensive existing literature on this class of TQPTs in semiconductors.

\textbf{Hamiltonian, TQCP, and phase diagram at finite temperatures:}
The semiconductor (\emph{e.g.,} InAs) system mentioned above
is mathematically described by the following Bardeen-Cooper-Schrieffer (BCS)-type
Hamiltonian:
\begin{equation}  \label{eq:polar_bulk_H}
H=(\eta \bm k^2-\mu)\tau_z + \Gamma \hat{\bm{S}}\cdot \bm\sigma+\frac{\alpha}{2}(\bm k\times \bm\sigma)\cdot\hat{\bm z}%
\tau_z+\Delta\tau_x
\end{equation}
where $\hat{\bm{S}}$ is a suitably chosen direction of the applied Zeeman spin splitting given by $\Gamma = 1/2 g\mu_B B$ with $g$ the effective
Land$\acute{e}$ g-factor,
$B$ the applied magnetic field and $\mu_B$ the Bohr magneton.
$H$ is written in terms of the $4$-component Nambu spinor $(u_\ua(\bm r),u_\da(\bm r),v_\da(\bm r),-v_\ua(\bm r))$,
and the Pauli matrices  $\sigma_{x,y,z},\tau_{x,y,z}$ act on the spin and particle-hole spaces, respectively.
 $H$ can describe a $2D$
system when $\bm k=(k_x,k_y)$ is a $2D$ vector with $\hat{\bm{S}}=\bm{\hat{z}}$, while
a $1D$ structure is described by choosing $\bm k=k_x$ with
 $\hat{\bm{S}}=\bm{\hat{x}}$.
Here, $\eta = 1/m^*$ with $m^*$ the effective mass of the
charge-carriers, $\mu$ is the chemical potential measured from the bottom of the top-most confinement induced band, the Zeeman splitting $\Gamma$ breaks the time reversal symmetry, $\alpha$ is the
Rashba spin-orbit coupling constant, and $\Delta
$ is an $s$-wave superconducting pair-potential proximity
induced in the semiconductor from an adjacent superconductor (\emph{e.g.,} Al).

The
Hamiltonian in Eq.~(\ref{eq:polar_bulk_H}) has recently been studied
extensively \cite{Sau,Index-Theorem,Alicea-Tunable,Roman,Oreg,Sau-PRB,Alicea-Network,Hassler,
Sau-Tewari-Universal,Akhmerov-Conductance,Mao,Sau-Interferometry,Sumanta-NJP}. A TQCP exists in
this system as the tuning parameter $\Gamma$ is varied through the critical
value $\Gamma = \Gamma_c = \sqrt{\Delta^2 +\mu^2}$ where the quantity $%
C_0=( \Delta^2 + \mu^2-\Gamma^2 )$ changes sign.
For $C_0 > 0$, the (low-$\Gamma$) state is an ordinary, non-topological
superconductor (NTS) with only perturbative effects from the Zeeman and spin-orbit
couplings. For $C_0 < 0$, however, the (high-$\Gamma$) state has
non-perturbative effects from $\alpha$, and can support zero-energy Majorana
fermion excitations localized at the defects of the pair-potential $\Delta$ \cite{Sau}. The parameters for the TQCP are in the
experimentally achievable range because, for a typical semiconductor wire say InAs or InSb, because of a large effective $g \sim 15-50$ a moderate
magnetic field $B\sim 0.5$ T corresponds to a Zeeman splitting $\Gamma \sim 2-8$ K. Noting that $\mu$ in Eq.~(\ref{eq:polar_bulk_H}) corresponds
to the chemical potential measured relative to the bottom of the top-most confinement band and $\Delta \sim 1-10$ K for an ordinary $s$-wave superconductor, a moderate $B\sim 0.5$ T should be sufficient to induce the topological phase transition in the semiconductor.
  For numerical calculations in this paper we have assumed $\mu =0$ (Fermi surface at the bottom of the top-most band), $\Delta = 0.5$ meV, $\alpha= 0.3$ meV, so that $\Gamma_c = 0.5$ meV. The existence of
Majorana fermions at defects, and also the fact that the high-$\Gamma$ state is isomorphic to a spin-less
$p_x+ip_y$ superconductor \cite{Alicea-Tunable,Zhang-Tewari,Sato-Fujimoto}, make the high-$\Gamma$ state a topological superconductor.
Interestingly, $\Delta$ remains non-zero and continuous across the TQCP \cite{Sau-PRB,Sumanta-NJP}, so the
NTS and TS states break exactly the same symmetries, namely, gauge
and time-reversal. As a result, no macroscopic local order parameter can
differentiate between the NTS and TS states, and they \emph{cannot} be distinguished
by any known bulk measurements.

 For our present purposes note that the topological critical point $%
\Gamma_c$ is marked by the single-particle minimum excitation gap $E_0$
vanishing as a function of the Zeeman splitting. This can be seen by
diagonalizing the Hamiltonian in Eq.~(\ref{eq:polar_bulk_H}) to obtain the
lower-branch of the quasiparticle excitation spectrum,
\begin{equation}
E_k^2=\Delta^2+\tilde{\epsilon}^2+r_{k}^{2}-2 \sqrt{\Gamma^2\Delta^2+\tilde{\epsilon}^2r_{k}^{2}},  \label{eq:spectrum}
\end{equation}
where $\tilde{\epsilon}=\eta k^2-\mu$ and $r_{k}^{2}=\Gamma^2+\alpha^2 k^2$. For $\Gamma$ near $\Gamma_c$, the
minimum of $E_k$ is at $k=0$, and setting $k=0$ in Eq.~(\ref{eq:spectrum})
we find the minimum quasiparticle gap $E_0$ given by,
\begin{equation}
E_0=|\Gamma-\sqrt{\Delta^2+\mu^2}|.
\label{eq:E0}
\end{equation}
$E_0$ vanishes exactly at $C_0=0$, which marks the TQCP separating the NTS
and TS states. Note that $E_0$ is finite and positive for both $C_0 > 0$
(NTS state) and $C_0 < 0$ (TS state).

It is important to note that the system exactly at the zero-temperature TQCP
($T=0, \Gamma=\Gamma_c$) can be
thought of as an $s$-wave superconductor ($\Delta$ is finite and
continuous at $\Gamma=\Gamma_c$). This bulk $s$-wave superconductor, however,
 coexists with nodal fermions at $k=0$.
Since on both sides of $\Gamma=\Gamma_c$ the
ground states are fully gapped, we can construct the  $(T-\Gamma)$ phase diagram
for this TQCP (Fig.~\ref{fig:phase_diagram}b) by drawing two crossover curves marking $k_BT \sim E_0(\Gamma)$
 on both sides of $\Gamma_c$.  Note that since $\Delta$ is non-zero everywhere on the phase diagram,
the finite-temperature crossover curves can only be justified, as we will show below (see the discussion following Eq.~\ref{scaling1}), on the basis
of some measurable quantities showing a pronounced change across these curves \cite{Sondhi-RMP}.
Such identification of crossover curves
opens a finite regime in the phase diagram which
we can associate with the quantum critical regime of this Zeeman-tuned TQCP.
The QC regime (and the crossover curves) can  also be understood \cite{Sachdev-Keimer} as the regime in which the
system, with increasing length scales, encounters the thermal length scale ($%
\beta=(k_BT)^{-1}$) \emph{before} it encounters the zero-temperature correlation length $\xi$
($\beta < \xi \sim (\Gamma-\Gamma_c)^{-1}$), which
diverges as $\Gamma$ approaches $\Gamma_c$.

\textbf{Bulk measurement (AC response):}
We now ask what sort of bulk measurements can access the nodal fermion
spectrum and reveal the underlying TQCP at $T=0$. At first
glance it may appear that a simple DC resistance measurement
may suffice, because, as a function of increasing $\Gamma$ at low $T$,
one should first see negligible resistance (NTS state with a gapped
spectrum) followed by nonzero resistance (QC regime with nodal
fermions) and finally again negligible resistance (TS state with
re-entrant gapped spectrum). This idea in practice would not work, however, because in
the presence of a DC voltage, the superconducting condensate will short the
 current out even in the QC regime, thus producing
negligible resistance everywhere in the phase diagram with increasing $\Gamma$.

\begin{figure}[tbp]
\centering
\includegraphics[scale=0.5,angle=0]{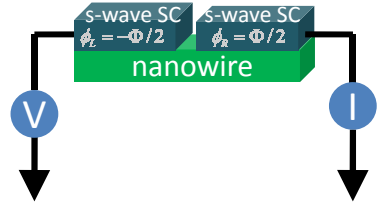}
\caption{Nanowire geometry for identifying the topological phase transition with increasing Zeeman splitting $\Gamma$.
The nanowire (shown in green) is contacted by two superconducting leads (blue) which produce the proximity effect.
The leads are placed with a finite potential difference $V$ and the current $I$ is measured. All quantities $V(%
\protect\omega)$, $I(\protect\omega)$ and $\Phi(\protect\omega)$ are
frequency-dependent. The josephson phase is $\Phi(\protect\omega)=-i V(\protect\omega)/\protect\omega$.}
\label{fig:geometry}
\end{figure}

We consider an alternative route to identifying the TQCP: an AC measurement.
For the sake of definiteness, we consider below (Fig.~\ref{fig:geometry}) the AC conductivity across a 1D nanowire
 contacted by $s$-wave superconducting leads which produces the proximity effect.
For AC conductivity measurements, depending on the frequency of the applied voltage,
there will be significant excitation of the nodal fermions only when the
energy scales corresponding to the frequency or the temperature become of the order of the single particle gap
$E_0$. The excited nodal fermions contribute to the dissipative (real)
part of the complex AC conductivity, while the superconducting condensate
 contributes to the inductive (imaginary) part. Thus, as a function of
increasing $\Gamma$ starting from deep in the gapped NTS state ($%
\Gamma=0$), one should first observe negligible, followed by non-zero, and then
re-entrant negligible dissipative response in AC conductivities. Such a behavior as a function of $\Gamma$ marks a
cross-over across the dashed curves in Fig.~\ref{fig:phase_diagram}b and in
turn reveals the underlying zero-temperature TQCP.

For quasi-2D and 3D systems
such a
dissipative AC response of the conductivity due to nodal fermions can be tracked by electromagnetic
absorption experiments \cite{Hosseini}. In the present case of a $1D$ semiconducting nanowire
proximity-coupled to bulk $s$-wave superconductors, such absorption
experiments can be difficult, and another quantity is needed which
nevertheless is still given by the real part of the AC conductivity
suitably defined. Below we show that the AC Josephson impedance serves this
purpose and it can be easily measured across a 1D semiconducting nanowire
contacted by $s$-wave superconducting leads which also produce the proximity effect (Fig.~\ref{fig:geometry}).

\textbf{AC Josephson impedance:}
The simplest AC response function that can be measured in a 1D nanowire in
proximity contact with superconductivity is the Josephson impedance.  The
linear response of the measured current $I(t)$ to a sinusoidal voltage $V(t)=V(\omega)e^{i\omega t}$
is written as $I(t)=(\chi_2(\omega)+i\chi_1(\omega))V(t)$ where $(\chi_2(\omega)+i\chi_1(\omega))^{-1}$
is the Josephson impedance of the junction \cite{mccumber} in Fig.~\ref{fig:geometry}. In the geometry of
 Fig.~\ref{fig:geometry}, when $\chi_2=0$, the junction behaves like a conventional
non-dissipative Josephson junction. However, when $\chi_2$ becomes non-zero, the voltage and current cease to be completely
orthogonal to each other and a finite amount of power ($\int dt I(t) V(t)$) is dissipated in the junction.
For the sake of brevity, below we will refer to $\chi_2$ simply as the dissipative susceptibility
keeping in mind that in reality this is the real part of the inverse impedance function.

We assume that the time-dependent voltages at the superconducting leads
are given by $V_L(x,t)=-V(t)/2$ on
the left lead and $V_R(x,t)=V(t)/2$ on the right lead. Correspondingly, the
superconducting phases on the right and the left leads are  $%
\phi_R(t)=-\phi_L(t)=\Phi(t)/2$, where $\Phi(t)$ is the total
time-dependent phase difference  between the leads. As is well known, $V(t)$
and the phase variation $\Phi(t)$ are related by the Josephson relation $%
V(t)=\dot{\Phi}(t)/2$.

The time-dependent BCS Hamiltonian describing the nanowire is given by
\begin{eqnarray}
H_{1} &=&\int dx\psi ^{\dagger }[-\partial _{x}^{2}+V(x,t)-\mu +\Gamma
\sigma _{x}-i\alpha \sigma _{y}\partial _{x}]\psi   \notag \\
&+&\int dx[\Delta (x)e^{i\phi (x,t)}\psi _{\uparrow }^{\dagger }\psi
_{\downarrow }^{\dagger }+h.c]  \label{eq:HBCS}
\end{eqnarray}%
where the pairing field $\Delta (x)$ is proximity-induced and therefore is
assumed to be non-zero only in the parts of the nanowire in direct contact
with the superconducting leads. $\psi ^{\dagger }=\psi _{s}^{\dagger }(x,t)$
creates a fermion in spin state $s=\uparrow ,\downarrow $ and spin-indices
are implicitly summed over. $V(x,t)$ is the voltage difference across the
wire. By applying the gauge transformation $\psi _{s}(x,t)\rightarrow
e^{i\Lambda (x,t)}\psi _{s}(x,t)$, $V(x,t)\rightarrow V(x,t)+\partial
_{t}\Lambda (x,t)$, $A(x,t)\rightarrow A(x,t)+\partial _{x}\Lambda (x,t)$
and $\phi (x,t)\rightarrow \phi (x,t)+2\Lambda (x,t)$ to the Hamiltonian in
Eq.~(\ref{eq:HBCS}), and by choosing $\Lambda (x,t)=-\phi (x,t)/2$, $H_{1}$
becomes,
\begin{eqnarray}
&&H_{1}=\int dx\psi ^{\dagger }[(-i\partial _{x}-A(x,t))^{2}-\mu +\Gamma
\sigma _{x}  \notag \\
&+&\alpha \sigma _{y}(-i\partial _{x}-A(x,t))]\psi +\Delta (x)[\psi
_{\uparrow }^{\dagger }\psi _{\downarrow }^{\dagger }+h.c]  \label{eq:HBCSA}
\end{eqnarray}%
where $A(x,t)=-\partial _{x}\phi (x,t)/2$. Since the voltage difference $%
V(x,t)$ drops smoothly across the junction with width $W$, we choose a phase
dependence $\phi (x,t)=\Phi (t)(2erf(x/W)-1)/2$ so that $\phi (x,t)$
increases from $-\Phi (t)/2$ for $x\ll W$ to $\Phi (t)/2$ at $x\gg W$.

Because of the spin-orbit coupling term in Eq.~(\ref{eq:HBCS}), the current
operator $J$ takes the following modified form:
\begin{equation}
J(x,t)=-i\left( \psi ^{\dagger }\partial _{x}\psi -\partial _{x}\psi
^{\dagger }\psi \right) -\alpha \psi ^{\dagger }\sigma _{x}\psi .
\label{current}
\end{equation}%
$H_{1}$ can then be rewritten as,
\begin{equation}
H_{1}(t)\approx H_{10}+\int dxJ(x,t)A(x,t),
\end{equation}%
(to linear order in $A$) where $H_{10}$ is $H_{1}$ at $A=0$.

The conductance of the wire is calculated from the current $J(x_{1},t)$ at a
position $x_{1}>W$ (outside the junction) in response to a perturbation $%
\int dxJ(x,t)A(x,t)$. In the limit of a small junction ($W\rightarrow 0$),
we can approximate the perturbation as $\Phi (t)J(0,t)/2$ and the measured
current as $J(0_{+},t)$. Therefore, choosing a time-dependent voltage $%
V(t)=V(\omega )e^{i\omega t}$ corresponding to a phase $\Phi (t)=V(\omega
)e^{i\omega t}/(i\omega )$ and using the fluctuation-dissipation theorem,
the real (dissipative) part of the current response function, $\chi
_{2}(\omega )=Re[\delta J(0_{+},\omega )/\delta V(\omega )]$ is given by
\begin{equation}
\chi _{2}(\omega )=\frac{1}{\omega }Im[\int_{0}^{\infty }dte^{-i\omega
t}\langle {[J(0,0),J(0,t)]}\rangle ].  \label{eq:chiformula}
\end{equation}%
We use Eq.~(\ref{eq:chiformula}) to calculate the dissipative susceptibility
$\chi _{2}$.

Substituting the operators $J(0,t)$ from Eq.~(\ref{current}) and calculating
the relevant matrix elements we obtain,
\begin{eqnarray}
&&\chi _{2}(\omega )=\frac{\pi }{2\omega }\sum_{n,m}\int \frac{dkdk^{\prime }%
}{(2\pi )^{2}}|\langle n,k|m,k^{\prime }\rangle |^{2}(k+k^{\prime })^{2}
\notag \\
&&\left[ f(E_{n,k})-f(E_{m,k^{\prime }})\right] \delta (E_{m,k^{\prime
}}-\omega -E_{n,k})  \label{eq:chi2}
\end{eqnarray}%
Here, $f(E)=(e^{E/T}+1)^{-1}$ is the Fermi occupation function, and $E_{n,k}$
and $|n,k\rangle $ are the Bogoliubov-de Gennes (BdG) eigenvalues and
eigenstates of $H_{1,0}$, respectively.
 This expression
is manifestly a bulk property which is real and positive only for positive
frequencies.

\begin{figure}
\centering
\includegraphics[scale=0.3,angle=0]{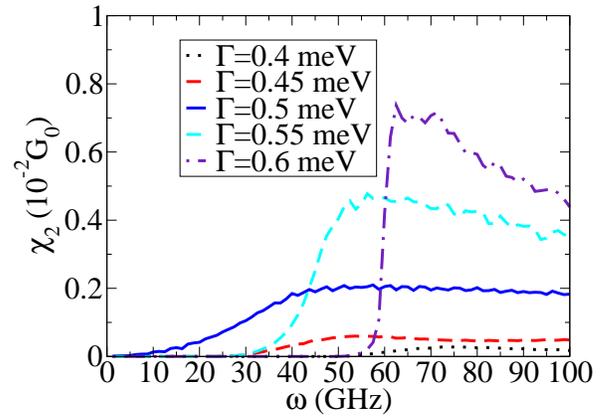}
\caption{Dissipative susceptibility $\chi_2$ ($G_0=2e^2/h$) as a function of frequency
$\omega$. If experimentally $\omega << T$ is satisfied, $\chi_2$ should be plotted with
the Zeeman splitting, see Fig.~\ref{fig3} and the discussion following it.
$\chi_2$ increases at lower-frequencies
as the Zeeman splitting $\Gamma$ is tuned through the TQCP ($\Gamma=\Gamma_c=0.5$ meV). For $\Gamma$ away
from $\Gamma_c$, in both the NTS and TS states the low-frequency ($\omega \ll E_0(\Gamma)$) dissipative response is negligible.
As the single-particle gap $E_0(\Gamma)$ closes at $k=0$, the threshold frequency for the dissipative susceptibility approaches zero at the
TQCP. Parameters are $\Delta=0.5$ meV, $\alpha=0.3$ meV, $\mu=0$ and $T=0.1$ meV corresponding to $1$ K. For reference
 we have used $1$ meV = $250$ GHz.}\label{fig2}
\end{figure}

The dissipative response $\chi_2$ vanishes sufficiently
far from the transition, both in the TS and NTS states, for $\omega$ smaller than the gap $E_0$.
To see this note that the integrand in Eq.~(\ref{eq:chi2}) is non-zero only when $E_{m,k^{\prime}}-E_{n,k}=\omega$ and
$E_{m,k^{\prime}}$ is empty together with $E_{n,k}$ being filled. At $T=0$, this can only happen when the single-particle gap
 $E_0$ is less than $\omega$.  The behavior of $\chi_2$ as a function of $\omega$ for fixed values of $\Gamma$ is shown in Fig.~\ref{fig2}.  As is clear from this figure, both the NTS and the TS states are non-dissipative for
frequencies smaller than a threshold set by $E_{0} (\Gamma)$.  As $\Gamma$ is tuned towards $\Gamma_c$ from either side, $E_0(\Gamma)$ decreases and the threshold
value of $\omega$ for the onset of dissipation decreases to zero at the TQCP.  Therefore, in the experiment suggested in Fig.~\ref{fig:geometry},  at sufficiently low $T,\omega \ll \Delta$ and away from the TQCP,
one would expect the measured current and the applied voltage to be out of phase by $\pi/2$ so that the
power dissipated is zero.  As one approaches the TQCP, at some value of $\Gamma$, $\omega$ will surpass
$E_0(\Gamma)$ and a component of the current will become in-phase with the applied voltage. This will lead to a finite power dissipation,
signaling the vicinity of the underlying TQCP.  The succession of behavior with increasing $\Gamma$ - non-dissipative,  followed by dissipative,
and then re-entrant non-dissipative response - of the Josephson current versus voltage is a clear signature of the underlying zero-temperature
TQCP. The re-entrant non-dissipative response for $\Gamma > \Gamma_c$ is also a clear signal of re-entrant \emph{high-Zeeman-field} superconductivity, which can only be topological in nature \cite{Sumanta-NJP}.

\textbf{Scaling of dissipative susceptibility:}
For analytical calculations of the scaling functions we first
need to derive the low-energy effective theory valid in the vicinity of the
TQCP. To do this we recall that the TQCP is given by the
minimum excitation gap $E_0$ vanishing as a  function of $\Gamma$. Near the
transition, only one pair of eigenstates of the BCS Hamiltonian in Eq.~(\ref%
{eq:polar_bulk_H}) vanishes near $k=0$. Therefore, near the transition, for
the low-energy effective theory we can `integrate out' the other pair of
Bogoliubov eigenstates and focus only on the lowest pair that vanishes at $\Gamma_c$.
The pair of eigenstates $n=1,2$, whose energies vanish linearly near $k\sim  0$, form a pair of
chiral Majorana fermion operators $\gamma_{n}(x)=\int dk e^{i k x}\sum_{s}[ u_{n,s}(k)\psi^*_{s}(k)+ v_{n,s}(k)\psi_{s}(k)] $. Here, $(u_{n,\ua}(k),u_{n,\da}(k),v_{n,\da}(k),-v_{n,\ua}(k))$ are the BdG eigenstates $n=1,2$ with eigenvalues $\pm E_k$ (see Eq.~\ref{eq:spectrum}) for $\Gamma=\Gamma_c$. The low-energy
effective action valid near the TQCP can then be written in terms of the Dirac fermions $\Psi^\dagger(x)=\gamma_1(x)+i\gamma_2(x)$ as, 
\begin{equation}
S=\int_0^\beta d\tau \int dx [\Psi^\dagger\partial_{\tau}\Psi +
i v(\Psi^\dagger\partial_x\Psi^\dagger +h.c)+\delta \Psi^\dagger\Psi],  \label{eq:action}
\end{equation}
where $\delta=(\Gamma-\Gamma_c)$ is the gap which takes the system away from
the phase transition and $v=\alpha\Delta^2/(\mu^2+\Delta^2)$ is a velocity determined by the spin-orbit
coupling constant $\alpha$. It follows that the dynamic critical exponent $z$%
, which relates the spatial and temporal correlation lengths $\xi$ and $%
\xi_{\tau}$ by $\xi=\xi_{\tau}^z$, is $1$. Since the energy gap $\delta$
vanishes linearly with $\Gamma$, $\xi_{\tau} \sim \delta^{-1}$ also diverges
linearly with the Zeeman coupling, and therefore the mean field correlation
length exponent $\nu=1$. A similar critical theory for this TQCP consisting of a
single species of gapless Fermion can be calculated in $D=2$ in an analogous manner.
In $D=1$ this is the
same as the critical theory of the Ising model in a transverse field, which in one dimension can be mapped into
Eq.~(\ref{eq:action}) by a non-local Jordan-Wigner transformation \cite{Sachdev}.

Note that the nodal quasiparticles constitute a quantum critical phase which is essentially a non-interacting
  gas of two species of chiral Majorana fermions defined by $\gamma_i^{\dagger}(k)=\gamma_i(-k), i=1,2$. The gaussian critical point
  implied in Eq.~(\ref{eq:action}) is stable against interactions because all four-fermion interaction terms can be shown to be
  irrelevant \cite{Sachdev}. Furthermore, the effective action in Eq.~(\ref{eq:action}) involves only one species of regular fermion $\Psi$ that
  becomes gapless at the critical point. This is a key difference between the Dirac spectrum found here and the analogous Dirac spectrum
  of the nodal quasiparticles in, say, $d_{x^2-y^2}$ superconductors or HgTe quantum wells \cite{molenkamp}
  where there are two species of gapless fermions corresponding to the spin degeneracy. Thus, our system avoids the fermion doubling theorem consequently giving rise to Majorana fermions and topological superconductivity whereas these other systems do not.

\begin{figure}
\centering
\includegraphics[scale=0.3,angle=0]{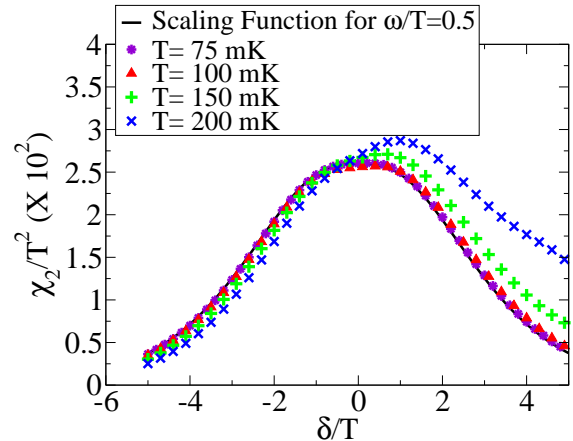}
\caption{
Dissipative susceptibility $\chi_2(T,\omega,\delta)$ shows scaling in the vicinity of the topological critical point $\delta=\Gamma-\Gamma_c=0$.
Appropriately scaled susceptibility data for $\omega/T=0.5$, calculated using Eq.~(\ref{eq:chi2}) for different values of
$\omega$ and $T$, coincide and collapse on the
scaling function (Eq.~\ref{scaling}) in the region near the critical point ($\delta\sim 0$). The slight asymmetry of the data
about $\delta=0$ is due to the asymmetry of the zero-temperature gap $E_0(\Gamma)$ about $\Gamma=\Gamma_c$ ($\delta=0$).
}\label{fig3}
\end{figure}

In the vicinity of the TQCP, with the effective critical theory in Eq.~(\ref{eq:action}),
the dissipative susceptibility  $\chi_2(\omega)$ for small $\omega$ and $\delta$ ($\omega,\delta$ much smaller than the gap at $k=k_F$) takes a
universal scaling form. This can be obtained by defining the rescaled variables, $%
\tilde{k}=k/T$, $\tilde{E}_k=E_k/T$, $\tilde{\omega}=\omega/T$ and $\tilde{\delta%
}=\delta/T$. The energy in Eq.~\ref{eq:spectrum} then takes the form $\tilde{E}=\sqrt{v^2\tilde{k}^2+%
\tilde{\delta}^2}$, while the matrix element $|\langle n,k|m,k^{\prime}\rangle|^{ 2}$
is invariant under the re-scaling. The dissipative susceptibility in the vicinity of the TQCP takes the
scaling form,
\begin{equation}
\chi_2=T^2 f(\omega/T,\delta/T)
\label{scaling}
\end{equation}
where the scaling function $f$ is given by,
\begin{align}
&f(x,y)=\sum_{m,n=\pm}\int \frac{d\tilde{k}(\tilde{E}_{n \tilde{k}}+x)}{8\pi x v^4
\tilde{k}^{\prime }}|\langle n,\tilde{k}|m,\tilde{k}^{\prime}\rangle^{ 2}(\tilde{k}+\tilde{k}^{\prime})|^{ 2}
\notag \\
&\left[\tanh\left(\frac{\tilde{E}_{n \tilde{k}}+x}{2}\right)-\tanh\left(\frac{\tilde{E}_{n,\tilde{k}}}{2}\right)\right].\label{scaling1}
\end{align}
Here $\tilde{E}_{\pm \tilde{k}}=\pm\sqrt{\tilde{k}^2+y^2}$ and $\tilde{k}^{\prime}$ in the integrand is implicitly given by
the equation $\tilde{E}_{m,\tilde{k}^{\prime}}=\omega + \tilde{E}_{n,\tilde{k}} $.

The existence of such a scaling function suggests that for $\omega\ll T$ $(\omega/T\rightarrow 0)$, $\chi_2/T^2=f(\omega/T,\delta/T)$ depends
 only on $\delta/T$. In this limit, the scaling function in Eq.~\ref{scaling1} becomes $f(0,y) \sim (1/\pi) y^2 e^{-y}$, which becomes appreciable only when the thermal energy scale $k_BT \sim \delta$.  This justifies our identification of
  $k_BT \sim E_0$ curves (near the TQCP, $E_0 = \Gamma-\Gamma_c=\delta$, see Eq.~\ref{eq:E0}) with the appropriate crossover curves in the finite temperature phase diagram in Fig.~\ref{fig:phase_diagram}b. For $\omega\gg T$, the argument $\omega/T$ in the scaling function $f$ approaches $\infty$ and must drop out. Thus, in this limit $\chi_2/T^2$ is a function only of $(\omega/T)(\delta/T)^{-1}=\omega/\delta$. This implies that near
   $T=0$, the dissipative susceptibility becomes appreciable only when $\omega$ approaches $\delta$ (and above). This is shown in Fig.~\ref{fig2}. In general, the validity of such a scaling function representation becomes clear from Fig.~\ref{fig3}, where
$\chi_2/T^2$ as calculated from Eq.~\ref{eq:chi2} is plotted for fixed $\omega/T$ and compared with the scaling function
Eq.~\ref{scaling1}. The collapse of the data for different $\omega$ and $T$ (but with the same fixed ratio $\omega/T$) on the scaling function (dashed curve) plotted as a function of $\delta/T$
is a clear and definitive experimental signature of the topological quantum critical point at $\Gamma=\Gamma_c$.

\textbf{Finite-$T$ crossover in supercurrent response:}
So far we have concentrated only on the dissipative part of the Josephson response of the nanowire bridge between the two superconducting
leads in Fig.~\ref{fig:geometry}. The dissipative part allows us to access the nodal quasiparticles across the QC regime.
The low-$T$ finite-frequency response indicates an underlying
gap collapse separating a fully-gapped $s$-wave superconductor at low $\Gamma$ from another fully-gapped $s$-wave superconductor at high $\Gamma$ that has identical broken symmetries (pair potential $\Delta$ remains the same). Although such a gap collapse at isolated points in the momentum space gives indications of an underlying
TQCP, the evidence nonetheless is still circumstantial.
In particular, no information about the specific topological nature of the critical point and the high-$\Gamma$ TS state can be derived from the behavior of the dissipative response across the QC regime. We
now consider an indicator of the topological character of the underlying zero-temperature critical point from the behavior of the corresponding supercurrent response
 across the QC regime at finite temperatures.

 The experimental set up in Fig.~\ref{fig:geometry} used to measure the finite-frequency dissipative response can be used to measure the supercurrent response as well. When the voltage $V$ is
time-dependent $V(t)=V(\omega)e^{i\omega t}$, the supercurrent response is given by the quantity $\chi_1(\omega)$.
This is expected to behave as $L_J/\omega$ throughout the phase diagram ($L_J$ is the effective inductance associated
with the Josephson junction), and therefore cannot distinguish between the NTS and TS states.
Let us therefore consider the limit $\omega\ll V(\omega)$ (where the linear response function $\chi_1(\omega)$ no longer
determines the supercurrent response) and in particular the case when $V(t)$ is time-independent. Even if $V(t)$ is time-independent the Josephson phase $\Phi$
 still linearly depends on time, $\Phi(t)=V t$. In this case, the Josephson current $I(t)$ through the wire bridge connecting
the superconducting leads in Fig.~\ref{fig:geometry} should oscillate sinusoidally with a frequency determined by the applied DC
voltage. Below we refer to this frequency as the AC Josephson frequency and derive its crossover behavior across the QC regime
at low and finite $T$.
At $T=0$, such an experiment has been proposed \cite{Roman,Oreg,Kitaev-Wire,Fu-Kane-Frac-Josh} to uniquely identify the topological
character of the nanowire TS state.
 We show that the AC Josephson frequency also shows a crossover across the $k_BT \sim E_0$ crossover line separating the QC regime and the TS
state in Fig.~\ref{fig:phase_diagram}. At this crossover a peak at a fractional frequency in AC Josephson effect becomes dominant and the conventional
Josephson frequency, characterizing the NTS state at small $\Gamma$, only makes a sub-dominant contribution. Tracking the AC Josephson
response across the QC regime on the same set up as that for the dissipative response can uniquely identify the underlying TQCP and the subsequent TS state at large $\Gamma$.

The current $I(t)$ across a nanowire junction between two superconducting leads with a voltage difference $V(t)$ is given by the
relation
$I(t)=\frac{1}{V(t)}\frac{\partial E_{tot}}{\partial_t}$,
where $E_{tot}$ is the total energy in the system. The energy of a Josephson junction can
be decomposed into two parts,
$E_{tot}=E_{junc}+E_{qp}$,
where $E_{junc}$ is the energy stored in the localized states around the junction while
$E_{qp}$ is the energy dissipated into the quasiparticles that propagate away from the junction.
The quasiparticle contribution to the current at low-frequency is dissipative and can be described by the dissipative
response function given earlier.
The energy stored in the junction, $E_{junc}$, is a function of both
the energy of the Andreev bound state $\epsilon_{ABS}(\Phi)$, which depends on the phase difference $\Phi$,
and the occupation number $n=0,1$ of the Andreev bound state. Specifically, shifting the
occupation number $n(t)$ of the junction switches the sign of $E_{junc}$ (because of particle-hole symmetry)
so that $E_{junc}(\Phi,n)=-(-1)^{n}\epsilon_{ABS}(\Phi)$. The usual Josephson supercurrent carried by the Andreev state in the
ground state (i.e. $n=0$) is dissipationless and
is given by $I_{SC}(\Phi)=-\frac{\partial \epsilon_{ABS}(\Phi)}{\partial\Phi}$, while the true current
through the junction $I(\Phi,n)=\frac{\partial E_{junc}(\Phi,n)}{\partial \Phi}$ has an additional factor of $(-1)^n$ \cite{Fu-Kane-Frac-Josh}.
For a fixed DC voltage, $\Phi(t)=V t$ and therefore
\begin{equation}
I(t)=I(\Phi(t),n(t))=(-1)^{n(t)}I_{SC}(V t).\label{In}
\end{equation}
In the NTS state of Eq.~(\ref{eq:polar_bulk_H}) the numerical results for $\epsilon_{ABS}$ can be fit by,
\begin{equation}
\epsilon_{ABS}(\Phi)= E_0\sqrt{1+D \cos{\Phi}},
\end{equation}
while in the TS state it crosses zero energy and can be approximated by
\begin{equation}
\epsilon_{ABS}(\Phi)=E_0 \sqrt{D}\cos{\frac{\Phi}{2}},
\end{equation}
where $D<1$ is the effective interface transparency \cite{Sengupta}.
 While these spectra (together with the particle-hole symmetric partners)
look similar, particularly in the regime $D\rightarrow 1$, they are
fundamentally different in terms of fermion parity.
In particular, for a fixed fermion number $n=0$ in the NTS state the energy of the junction $E_{junc}(\Phi,n=0)$  remains
negative for all values of the phase $\Phi$, while in the topological state the state with a fixed fermion parity $n$ crosses
 zero energy.
The AC Josephson current in the presence of a DC voltage qualitatively
distinguishes between the NTS and TS states.
In the NTS state we get a ground-state (i.e. $n=0$) current,
\begin{equation}
I(n(t)=0,t)=\Delta \frac{D\sin{V t}}{\sqrt{1+D \cos{V t}}},
\end{equation}
which has harmonics only at multiples of  $V$.
In contrast the current in the topological phase is given by the relation
\begin{equation}
I(n(t)=0,t)=\Delta \sqrt{D}\sin{\frac{V t}{2}}
\end{equation}
and has a frequency of $V/2$.

The above picture for the supercurrent in terms of Andreev bound states is only valid for biases that
are smaller than the bulk gap of the system i.e. $V\lesssim E_0$. For higher biases, $V\gtrsim E_0$, the
harmonically varying superconducting pairing term $\Delta e^{i V t}$ can excite a quasiparticle
out of the Andreev bound state into the quasiparticle gap making the Andreev bound state ill-defined.
Therefore the definition of the Josephson effect is valid for voltages significantly smaller than the gap.

\begin{figure}
\centering
\includegraphics[scale=0.3,angle=0]{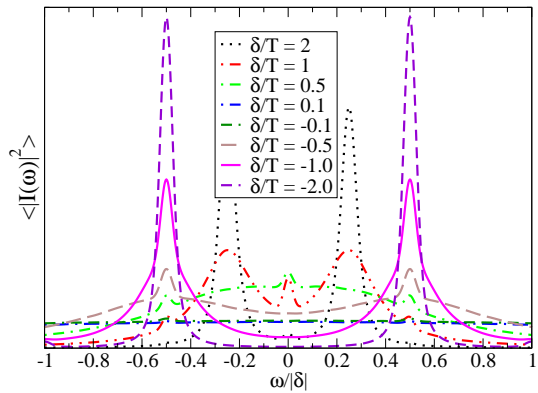}
\caption{
Spectral decomposition of current $\langle{|I(\omega)|^2}\rangle$ as a function of rescaled
frequency $\omega/|\delta|$ where $\delta=\Gamma-\Gamma_c$
at several values of $\delta$ characterized by $\delta/T$. The applied DC voltage is
taken to be $V=\delta/2$ so that the AC Josephson current in the non-topological phase
(i.e. $\delta<0$) shows a peak at $\omega=\delta/2$ characteristic of the conventional
AC Josephson effect. At and above $\delta/T \sim 1$ (i.e. the crossover curve $k_BT \sim E_0$ for $\delta>0$) a peak appears at
$\omega=\delta/4$ which signals the appearance of the fractional Josephson effect and an underlying $T=0$ TQCP at $\delta=0$.
}
\label{fig:frac}
\end{figure}

For voltages that are smaller than the gap (i.e. $V\lesssim E_0$), the Josephson current has a sign that is
determined by the occupation of the Andreev bound state, namely
\begin{equation}
I(t)=(-1)^{n(t)}I_{SC}(V t),
\end{equation}
where $n(t)$ is the occupation of the Andreev bound state. Here we assume $n(t)$ to change instantaneously
between the values $0$ and $1$ and \emph{vice versa}. Fermion parity conservation requires that such a change be accompanied by
the emission of a quasiparticle which costs energy $\epsilon_{qp}>E_0$. Thus the transition from the state $n=0$ to $n=1$ requires
an energy of $\epsilon_{qp}+\epsilon_{ABS}(\Phi)$ while the reverse transition requires an energy  $\epsilon_{qp}-\epsilon_{ABS}(\Phi)$. At zero temperature such excitations are forbidden and the Josephson effect operates without fluctuations of $n(t)$ as
discussed before. At finite $T$, the energy required for the above transitions is provided by thermal fluctuations either in the
form of phonons or quasiparticles. In the following we refer to such excitations as phonons even though,
the results will apply to more general excitations. For simplicity, we assume that a phonon above the required energy
threshold will flip $n(t)$ whenever it is energetically allowed. For temperatures $T\ll E_0$, the density of phonons
with energy $\epsilon$ incident at the junction is given by the Bose-Einstein distribution
$
n_{BE}(\epsilon)=\frac{1}{e^{\epsilon/T}-1}.
$
Assuming that the phonons move with the sound velocity, $v_k$, the rate of transitions at the junction for energy bigger than $\epsilon$ is given by
\begin{equation}
\rho_{phonon}(\epsilon)=\int_\epsilon^\infty d\epsilon'\frac{v_k(\epsilon') D(\epsilon')}{e^{\epsilon'/T}-1}\sim T e^{-\frac{\epsilon}{T}}
\end{equation}
for $E_0\gg T$. Here we have used that $D(\epsilon)=(\frac{d\epsilon(k)}{d k})^{-1}$ and $v_k=\frac{d\epsilon(k)}{d k}$.
 As expected the rate has the dimensions of energy.
The corresponding flip rates can then be written as
\begin{align}
&P(n(t)=0\rightarrow 1)=T e^{-(E_0+\epsilon_{ABS}(\Phi(t)))/T}\\
&P(n(t)=1\rightarrow 0)=T e^{-(E_0-\epsilon_{ABS}(\Phi(t)))/T}.
\end{align}

In the low temperature limit $(T < E_0)$ one can apply a voltage $V$ such that $E_0(1-\sqrt{D})> V > T e^{-E_0/T}$. In this case the Josephson
oscillation frequency is much bigger than the flip rate for $n(t)$. For runs of $I(t)$
where $n(t)$ changes only after many periods of oscillation of $I_{SC}(t)$, the Fourier
transform should show a pronounced peak at $\omega=V$ for the regular Josephson effect (on the NTS
side) and $\omega=V/2$ for the fractional Josephson effect (on the TS side).
In the temperature range $T \gtrsim E_0$ one is restricted to low voltages, $V < T e^{-E_0/T}$, since $V$ is
bounded by $E_0$. In this case the dynamics of the quasiparticle
state $n(t)$ is rapid compared to the phase and one can assume that $n(t)$ is in local equilibrium, so
that the resulting current $I(t)$ will be a $2\pi$-periodic function of the phase $\Phi=V t$. To show the crossover in the frequency dependence of the Josephson current across $T \sim E_0$, we calculate $I(\omega)$ as the Fourier transform of $I(t)$ in Eq.~(14) and calculate the average $\langle{|I(\omega)|^2}\rangle$ where the average is taken with respect to random realizations of $n(t)$ according to Eqs.~(16,17). In Fig.~\ref{fig:frac} we show the evolution of the frequency dependence of the AC Josephson response at fixed low $T$ for various
values of $\delta/T$ (a horizontal cut across the QC regime in Fig.~\ref{fig:phase_diagram}b). The finite-$T$ Josephson response shows a pronounced crossover across the $k_BT \sim E_0$ curve separating the QC regime from the TS state at large $\Gamma$. At this crossover, frequency peaks
 indicating an underlying zero-$T$ Josephson period-doubling transition starts making a dominant contribution and the usual Josephson frequency becomes sub-dominant. Such a crossover in the superfluid response can serve as an unambiguous marker of an underlying zero-$T$ TQCP in the semiconductor heterostructure.

\textbf{Summary and Conclusion:}
TQCPs separate two macroscopic ground states which have the same symmetries and hence cannot be distinguished by
an order parameter. Consequently, the topological and the non-topological states on the two sides of a TQCP cannot in general
be distinguished by any bulk measurement, \emph{e.g.,} no thermodynamic quantity diverges at the TQCP. To solve this problem, we propose the use of specific aspects of the topological phase transition itself as an identifier.
 We consider the TQCP in a spin-orbit coupled semiconductor (\emph{e.g.,} InAs) thin film or
nanowire on which $s$-wave superconductivity is proximity induced. The TQCP in this case is tuned by an external Zeeman splitting $\Gamma$, and
for large $\Gamma$ the ground state of the system is a topological superconductor.
We ask if straightforward bulk transport measurements can help
identify the emergence of the TS state with increasing values of $\Gamma$. We show that this is indeed possible and establish that the finite-$T$ AC Josephson impedance along with the supercurrent response of the semiconductor can access the topological \emph{quantum critical} regime which precedes the TS state in the finite-$T$ phase diagram (Fig.~\ref{fig:phase_diagram}).

We do this by first identifying the entire QC regime with a bulk $s$-wave superconductor coexisting with gapless nodal fermions at $k=0$.
Since both the non-topological superconducting state at low $\Gamma$ and the TS state at
high $\Gamma$ are fully gapped, both these states are devoid of quasiparticles at sufficiently low $T, \omega \ll E_0$, where $E_0(\Gamma)$ is the
Zeeman-tunable single-particle energy gap.  The real part of the inverse AC Josephson impedance ($\chi_2$), which gives the dissipative response of a nanowire contacted by two superconducting leads (Fig.~\ref{fig:geometry}), is therefore negligible away from the TQCP for $\omega, T$ less than $E_0$ in both the NTS and TS states. As the TQCP is approached with increasing $\Gamma$,
the Josephson impedance for a given frequency picks up as the frequency becomes comparable with the decreasing values of the energy gap. In the low temperature QC regime in the finite-$T$ phase diagram, the impedance follows a scaling function involving $T, \omega,$ and $\delta=\Gamma-\Gamma_c$, which can be experimentally verified to help reveal the underlying quantum critical point. When the Josephson impedance for a given frequency decreases again with increasing values of $\Gamma$ past $\Gamma_c$, it indicates the emergence of the TS state on the high-Zeeman-field side of the critical point. The finite-$T$ superfluid response in the set up of Fig.~\ref{fig:geometry} also shows a pronounced crossover at the $k_BT=E_0$ curve
on the large-$\Gamma$ side of the phase-diagram (Fig.~\ref{fig:phase_diagram}b). At this crossover, the usual Josephson current frequency in response to a DC voltage becomes sub-dominant and pronounced peaks at a fractional frequency dominates the spectral decomposition of the Josephson current (Fig.~\ref{fig:frac}). Such a finite-$T$ crossover in the superfluid response can be observed in the same set up as in Fig.~\ref{fig:geometry} and can serve as an unambiguous indicator of the underlying $T=0$ TQCP in the semiconductor heterostructure. Our work demonstrates that a bulk experimental characterization of TQCP may indeed be possible although it may require a careful analysis of specific properties of the topological phase transition as we have carried out here for the TS state in semiconductor heterostructure systems with the real significance of our work lying in the fact that no TQCP has yet been clearly identified experimentally in any system.

\textbf{Acknowledgements:} This work is supported by DARPA-MTO, NSF, DARPA-YFA, DARPA-QuEST, JQI-NSF-PFC, and Microsoft-Q. We thank L. P. Kouwenhoven and C. M. Marcus for fruitful discussions.

\end{document}